\documentclass[10pt,letterpaper]{article}
\usepackage{opex3}
\usepackage{amsmath,amssymb,graphicx}

\begin{document}

\def\mum{\,$\mu${\rm m}}
\newcommand{\vdag}{(v)^\dagger}
\newcommand{\um}{$\mu$m}
\newcommand\arcsec{\mbox{$^{\prime\prime}$}}%
\newcommand{\unit}[1]{\ensuremath \;\mathrm{#1}}
\newcommand{\gammaray}{$\gamma$-ray\xspace}
\def\ni{\noindent}
\def\pomega{\tilde{\omega}}
\providecommand{\nn}{\nonumber}
\providecommand{\la}{\lambda}
\providecommand{\e}[1]{\ensuremath{\times 10^{#1}}}
\providecommand{\tr}[1]{\textrm{#1}}

\title{A computationally efficient autoregressive method for generating phase screens with frozen flow and turbulence in optical simulations}

\author{Srikar Srinath$^{1*}$, Lisa A. Poyneer$^2$, Alexander R. Rudy$^1$ and S. Mark Ammons$^2$}
\address{
   $^1$Dept. of Astronomy \& Astrophysics, University of California Santa Cruz, Santa Cruz 95064, USA \\
   $^2$Lawrence Livermore National Laboratory, Livermore, CA 94450, USA}

\email{$^*$ssrinath@ucsc.edu} 


\begin{abstract} We present a sample-based, autoregressive (AR) method for the generation and time evolution of atmospheric phase screens that is computationally efficient and uses a single parameter per Fourier mode to vary the power contained in the frozen flow and stochastic components. We address limitations of Fourier-based methods such as screen periodicity and low spatial frequency power content. Comparisons of adaptive optics (AO) simulator performance when fed AR phase screens and translating phase screens reveal significantly elevated residual closed-loop temporal power for small increases in added stochastic content at each time step, thus displaying the importance of properly modeling atmospheric ``boiling''. We present preliminary evidence that our model fits to AO telemetry are better reflections of real conditions than the pure frozen flow assumption.\end{abstract}



\section{Introduction}
\label{Intro}

End-to-end simulation of the next generation of telescopes, their adaptive optics systems, instrumentation, and control systems is increasingly a necessity given their high cost and complexity. For existing systems with planned upgrades or changes, minimizing disruption by conducting rigorous computer or lab testbed simulations \cite{Rudy14} is essential. Simulations of the entire optical pipeline and control algorithms have been employed with great success in programs such as the Gemini Planet Imager (GPI) \cite{Poyneer06}. Systems under construction (the Large Synoptic Survey Telescope (LSST) \cite{Connolly10}) or in the early design stage (the Thirty Meter Telescope \cite{Wang12}) have devoted significant fractions of their time and monetary budgets to producing detailed error budgets. Systems like LSST, which are particularly sensitive to   
atmospheric variability \cite{Jee11}, need atmosphere models that closely match real conditions while not extending their already significant computational requirements.

Underpinning all simulations is the atmosphere model through which wavefronts from artificial sources propagate and accumulate phase errors that a simulation in turn propagates through a telescope's optics to observe the effect on the system's point spread function (PSF). Systems with active \cite{Connolly10} or Adaptive Optics (AO) \cite{Wang12} then make their best effort to correct for these phase errors. One of the more prevalent methods used to mimic wind in simulations is by invoking Taylor's frozen flow hypothesis: generating large phase screens of zero-mean, Gaussian white noise in the time domain, scaling them according to the desired statistics ({\em e.g.} Kolmogorov) in Fourier space \cite{Mcglam76}, and translating them across the aperture at each time step. One phase screen is generated for each wind layer with dimensions based on exposure time, desired screen resolution, wind speed and direction in that layer, and considerations like making the array square and close to a power of 2 to apply a Fast Fourier Transform (FFT). While this method has proven its utility thus far, the advent of large-aperture telescopes \cite{Wang12} or ones with wide fields of view means \cite{Connolly10} that simulations will and do rapidly run up against computation and storage limits, particularly for high-resolution or long-exposure runs. For example, the GPI Adaptive Optics Simulator (GPIAOS) is limited to a 4-second maximum exposure because of memory limits and the LSST simulation team has to generate phase screens that are tens of gigabytes per wind layer for their runs \cite{Peterson15}.

There are two major recent thrusts in phase screen generation research. 1) Solutions that preclude the need to generate large phase screens exist or are being developed, such as extending aperture-sized screens by a few columns at each time step \cite{Assemat06, Fried08}. However, they do not address atmospheric ``boiling'' and changing the atmospheric turbulence model used is not trivial. 2) There are also efforts to increase the realism of simulations by generating and evolving turbulence at multiple scales \cite{Beghi13, Vorontsov08} or evolving atmosphere parameters, like phase coherence length \cite{Assemat06}. In this paper we present an autoregressive (AR) method that encompasses aspects of both research direction: it is a computationally efficient method that can be used to simulate frozen flow for long-exposure simulations or for wide-aperture systems; it accommodates a wide variety of turbulence models (Kolmogorov, von Karman, bespoke); it uses one parameter ($\alpha$) to vary stochastic content and another ($P$) to set the turbulence model parameters, both of which can be varied in time, as outlined in Section \ref{sec:method}. The AR method starts with the translating screen generation method described in the previous paragraph \cite{Johan94}, advances phase in the Fourier domain to simulate wind and adds in appropriately scaled noise at each time step to inject random variations to simulate ``boiling.'' It provides a path around computational resource limitations and is simple to implement, especially when compared to the complexity of modal methods \cite{Harding03}. The AR method is compatible with any sample-based phase screen generation method that operates in the Fourier domain. Both major shortcomings of FFT-based methods cited as motivation for current phase screen work \cite{Assemat06}, finite exposure times and static atmosphere parameters, are addressed by the AR method.

A method using a Markov process with elements similar to the one we propose is described in \cite{Glind93}. Phase translation to simulate wind occurs in the Fourier domain and scaled, decorrelated phase is added to an instantiation of a phase screen. However, the decorrelation coefficient (analogous to our autoregression coefficient) increases with spatial frequency, something that we find does not correspond to measured telemetry as we outline in Section \ref{sec:motivation}. In addition, the method is subject to periodicity, a shortcoming common to all Fourier-based methods, which results in phase screen sizes increasing with simulation length. Periodicity is an issue we explicitly address below and in Section \ref{ssec:periodicity}. There is no change in phase screen size as simulation time increases using the autoregressive method described in this paper. The authors also do not make it explicitly clear whether they add decorrelated phase at each time step in a simulation, while we outline our algorithm in an easily replicable manner with detailed analysis of its memory and computation requirements. 

Two additional criticisms of sample-based Fourier methods are the periodicity of generated screens, which wrap around the domain, and the fidelity of their power spectra to theory, in particular at low and high spatial frequencies. Algorithms to eliminate periodicity by extending the physical size of the screen \cite{Vorontsov08} exist and can be applied to the AR method, but, as we demonstrate in Section \ref{ssec:periodicity}, the process of adding stochastic content with the desired power spectrum at each time step effectively results in the renewal of phase after a time interval that scales with the magnitude of the autoregressive scaling coefficient used. 

Efforts to address the power spectrum problem have ranged from hybrid, covariance-based, Fourier-modal representations of phase \cite{Welsh97} to modifications of the sample-based Fourier methods \cite{Lane92, Johan94}. The sample-based methods have tended to concentrate on low spatial frequency power because spectral roll-off at the high-frequency end means that the low-frequency power mismatch is the more significant problem \cite{Sedmak04}. A hybrid method \cite{Welsh97}, while producing screens whose power content agrees well with theory, still invokes frozen flow, suffers from periodicity and is order $O(n^2)$ in complexity (where $n$ is the number of points in the screen). Methods that add subharmonics to boost low-frequency power are efficient augmentations to the AR method as long as the power spectrum scaling coefficient (denoted $\beta$ in Section \ref{sec:method}) is appropriately adjusted to conserve total power. The Markov process screen generation method described above includes an analysis of how to add a computationally reasonable set of subharmonics that will yield a spatial power spectrum which conforms more closely to theory \cite{Glind93}. However, we are in the process of analyzing telemetry gathered from instruments such as GPI and the ShaneAO Adaptive Optics system \cite{Gavel14} on the Shane 3-meter telescope at Lick Observatory and will exploit the capacity to modify power in each Fourier mode to match measured data. 

Section \ref{sec:method} describes the AR phase screen generation method in detail. Its reduced memory requirements, computation complexity that is at least comparable to (if not less complex) than commonly used existing phase screen generation methods and lack of periodicity are analyzed in Section \ref{sec:memory}. We test phase screen PSDs for fidelity to theory and illustrate the impact of adding small amounts of boiling to frozen flow phase on the GPI AO simulator, which incorporates large parts of the as-built system, in Section \ref{sec:validation}. Key metrics such as contrast ratio are reduced by factors of 2--20 in the central regions of a target bright star's PSF, {\em i.e.} in precisely the area the system is expected to find and take spectra of exoplanets. We analyze differences in power content of AR atmospheres and frozen flow screens. Finally, we provide an example of fitting collected telemetry with the AR model in Section \ref{sec:motivation}. This section serves to provide justifications for our assumptions on model parameters and to tie this work into the broader picture of our ongoing work on wind predictive AO. Wind predictive AO relies on Kalman filters generated from AR model fits to telemetry in real-time. The filter generator will be trained on more realistic phase screens that incorporate multiple wind layers and ``boiling'' parameters extracted from a large corpus of existing telemetry from GPI and the Shane telescope.

\section{Description of method}
\label{sec:method}  

The key differentiator between the AR method and others is the ability to tailor atmospheres and vary power in frozen flow and stochastic components at a modal level. Each Fourier mode has an associated complex autoregressive scaling parameter, $\alpha$ with $|\alpha|\leq 1$, which attenuates phase from the previous timestep. In addition, the phase of $\alpha$ encodes wind velocity and direction for a given layer \cite{Poyneer07}. Multiplying by an exponential in the Fourier domain translates each Fourier mode resulting in a frozen flow effect which wraps around the domain. Setting $|\alpha|=1$ produces pure frozen flow. 

The autoregressive method for a given wind layer starts at time $t=0$ by generating a phase array, $\phi_0$, in the commonly used manner outlined above which is slightly wider than the telescope primary pupil diameter to mitigate edge effects. White noise is scaled in Fourier space by a Kolmogorov power law, $P$, ({\em n.b.} $P$ can also follow von Karman statistics where outer scale is important or other turbulence models, even, as we foreshadow in Section \ref{sec:motivation}, fits to telemetry) to give it the appropriate power spectrum \cite{Johan94} as follows: 

\begin{equation}
P \propto \frac{2 \pi}{S} N_p r_0^{-5/6} \left(f_x^2 + f_y^2\right)^{-11/12}
\end{equation}

\ni where:
\vspace{-2mm}
\begin{itemize} \itemsep1pt \parskip0pt \parsep0pt
\item $S =$ screen diameter in meters
\item $N_p =$ number of pixels across screen (which is $N_p\times N_p$ in extent)
\item $r_0 =$ Fried's parameter \cite{Fried65} or phase coherence length in meters at a wavelength of 500 nm
\item $f_x,f_y$ = spatial frequency components (meter$^{-1}$). 
\end{itemize}

For each subsequent iteration, we treat the evolution of phase, $\phi$ (in radians), as an autoregressive process of order 1 in Fourier space, denoted AR(1) -- {\em i.e.} one where the phase at at any given time step is only dependent on the phase from the previous time step and added noise scaled by $P$ \cite{Percival93}. For Fourier-transformed (FT) phase from the previous timestep, $FT(\phi_{t-1}) = \widetilde{\phi}_{t-1}$, each Fourier mode ($k \propto N_p f_x,\ l \propto N_p f_y$) is scaled by $\alpha$. AR coefficient $\alpha$ is a complex array, $N_p \times N_p$ in extent, and a function of spatial frequency ($\alpha = \alpha(f_x, f_y)$) so each frequency can be individually modified. The magnitude of $\alpha$ determines how much past phase is attenuated at every time step. The phase of $\alpha$ (designated $\theta$ to avoid confusion with atmospheric phase) contains wind velocity and direction information. It translates $\phi_{t-1}$ in the Fourier domain\cite{Poyneer07} and is given by: 

\begin{equation}
\theta = -2\pi T \left(f_x v_x + f_y v_y\right)
\label{eqn:alphaphase}
\end{equation}

\ni where:
\vspace{-2mm}
\begin{itemize} \itemsep1pt \parskip0pt \parsep0pt
\item $v_x, v_y$ are wind velocity components (meters s$^{-1}$)
\item $T$ is the sampling interval (in seconds)
\end{itemize}

Appropriately scaled, Fourier-transformed, unit-variance, spatially white noise ($\widetilde{\omega}$, also an $N_p\times N_p$ array, $\sigma^2_{\omega}=1$) is added back in at each timestep to simulate the effect of turbulence. The white noise scale factor ($\beta = \sqrt{1-|\alpha^2|} P$) is a function of $\alpha$ and $P$ so that total power is conserved across phase screen time series as $\alpha$ varies -- something that we verify in Section \ref{sec:validation}. Any method used to boost low spatial frequency power would also require modifying $\beta$. The atmospheric phase in the Fourier domain at any timestep $t$ is:

\begin{equation}
\widetilde{\phi}_t = \alpha \widetilde{\phi}_{t-1} + \sqrt{1-|\alpha^2|} P \widetilde{\omega}_t
\label{eqn:areqn}
\end{equation}

\section{Memory and computational costs}
\label{sec:memory}

\begin{figure}[h!]
\includegraphics[width=1.0\columnwidth]{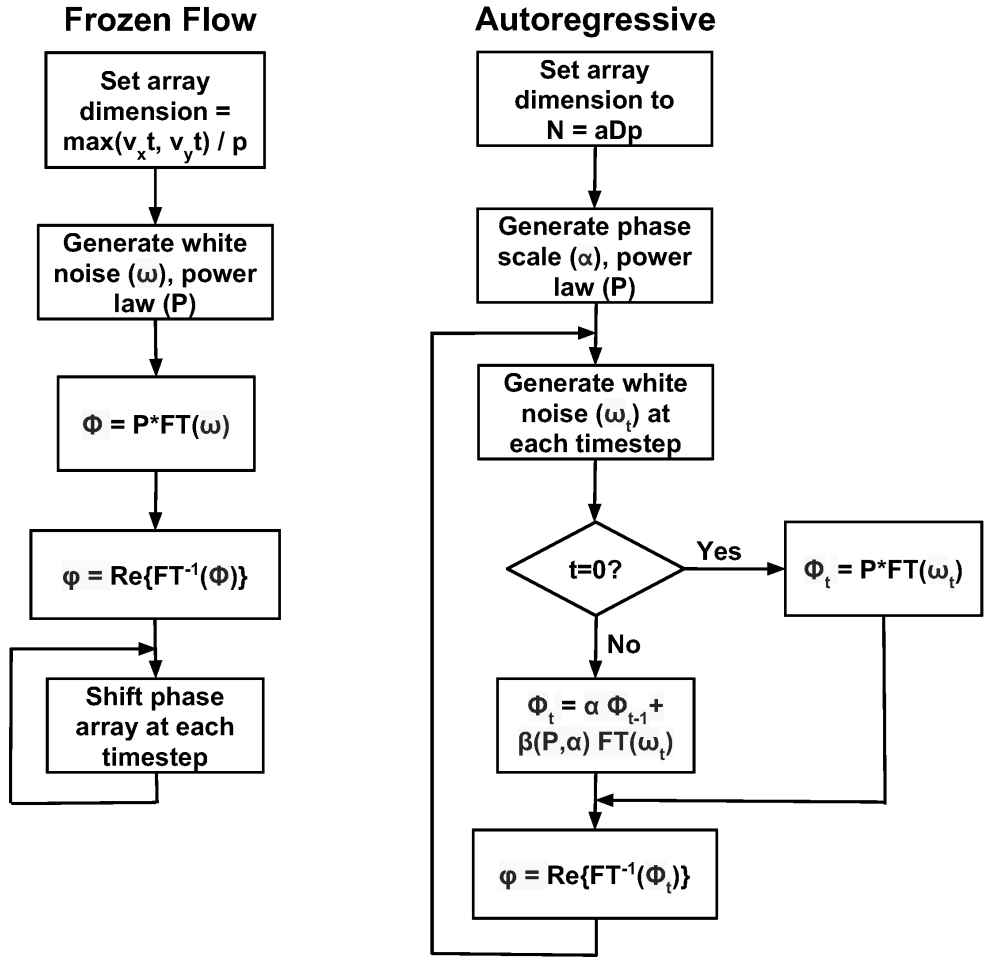}
\caption
   { \label{fig:flowchart} 
	Flowcharts comparing the traditional frozen flow atmosphere generation process (\textit{left}) and the autoregressive AR(1) process (\textit{right}) for a 1-layer atmosphere (the process is repeated for each wind layer). The seeming simplicity of the frozen flow flowchart belies its memory requirements. In the autoregressive method, $|\alpha| \leq 1$ and this determines by how much prior atmospheric phase is attenuated. The phase of $\alpha$ carries information about wind velocity and direction in the layer as seen in equation \ref{eqn:alphaphase}.}
\end{figure}

When using a FFT, spatial frequencies are indexed as Fourier modes, ($k, l$). Flowcharts comparing the traditional frozen flow method and this new autoregressive method are shown in Fig. \ref{fig:flowchart}.

For our implementation of this method we chose a computationally efficient DFT method, FFTW3 \cite{Frigo05}. Each complex Fourier transform is of order $5 N \log_2(N)$ floating point operations (flops), where $N= N_p^2$. FFTW3 has the advantage of delivering the same performance for values of N that are not powers of 2 as well, which is not the case for conventional implementations of the FFT algorithm (such as that bundled with the IDL programming language). As seen in Fig. \ref{fig:flowchart}, in the normal course of operation, each iteration involves generating an $N_p \times N_p$ array of white, Gaussian noise, followed by a scaling operation and a Fourier transform (FT)/inverse Fourier transform (IFT) pair for each wind layer. Other than the FT/IFT pair, the other operations are element-wise, 2-D array operations of order $O(N)$. Hence, the overall complexity of the AR algorithm is of order $O(N\log N)$. More realistic simulations will have to evolve $r_0$ and wind speeds in time, which means $P$ and $\alpha$ will have to be recalculated periodically in the course of a simulation. While this will increase computation time, the memory requirements and complexity will not change. 

The AR method's computational complexity is of the same order as the traditional frozen flow method ($O(N\log N)$), with a significantly smaller N, as shown later in this section. When comparing per iteration, sub-pixel shifts of a large array to simulate wind in any direction for the conventional method can be done in the Fourier domain on a sub-array around the aperture. This entails performing a FT/shift/IFT operation on an array at least as large as the AR phase array, so the AR method performs no worse, computationally speaking, per timestep. Both methods could improve performance by generating Hermitian white noise in the frequency domain \cite{Richards07}, which removes the need for an FFT per timestep. Covariance-based methods \cite{Harding03} can do no better than $O(N^2)$, with $N$ inflated by the requirement of creating a large, translating screen to mimic wind using a frozen flow model. 

Phase screen extrusion methods \cite{Assemat06, Fried08} which extend phase screens by a few columns, 
$N_{col}$, in the direction of wind flow compute a covariance matrix at each timestep which is a matrix multiplication operation involving a matrix of size $N_p \times (N_p \times N_{col})$ and a vector with $N_p \times N_{col}$ elements. This is an $O(N)$ operation (where $N= N_p^2$ as noted above)  -- more efficient than the AR method. 

If a simulation is tracking more than one atmosphere layer, phase screen generation and evolution of each layer is handled separately in the Fourier domain before a linear combination of the real part of their IFTs is used as the new wavefront in a given simulation timestep. Memory requirements for a multiple wind layer simulation are:

\begin{equation}
M = 2b\ (n+1) N
\label{eqn:memeqn}
\end{equation} 

\ni where:
\vspace{-2mm}
\begin{itemize} \itemsep1pt \parskip0pt \parsep0pt
\item $n$ is the number of wind layers (+1 for the cumulative phase that is actually fed into simulation)
\item $N = N_p^2$, number of points and $N_p= aD/p $, the pixel width of the screen. $a$ is a scale factor on the order of unity used to scale up screen size to the nearest size convenient for FFTs, reduce periodicity or increase low-order spatial power. $D$ is entrance pupil diameter in meters and $p$ is the spatial sampling scale or resolution of the phase screen (meters pixel$^{-1}$)
\item $b$ is the number of bytes per phase array element (varies depending on the precision desired -- typically 4 or 8) and the factor of 2 arises because phase in the Fourier domain is a complex number
\end{itemize}

For example, phase screens used with the GPIAO simulator are 48 subapertures across and each subaperture is 8 pixels wide, resulting in $384 \times 384$ pixel screens for each time step. The Gemini telescope's primary mirror is 7.7 meters in diameter with 43 subapertures across it. A complex phase screen Fourier array for a single wind layer (regardless of wind speed, direction and exposure time/simulation length) occupies $8 \times 384^2 = 1.18$ MB for a given time step in the IDL-based GPIAOS. In contrast, a translating frozen flow phase screen with a typical wind velocity measured on Cerro Pachon of $\sim 10$ m/s East-southeast \cite{Tokovinin06} requires a large value of $N$ (Equation \ref{eqn:memeqn}) to account for the wind vector and simulation length. Since the screen has to be square and $N_p$ can only be a limited set of values ({\em e.g.} powers of 2 form one subset of allowable $N_p$ values), a 1-second simulation requires a $1536 \times 1536$ pixel screen (18.9 MB) per wind layer while a 4-second run demands a $4096 \times 4096$ pixel screen (134 MB) per layer. In addition, the entire screen for a wind layer has to be carried in memory (though memory mapping can solve the problem of RAM capacity with the trade-off of slowing down simulation time with hard drive accesses) and manipulated at each timestep. Hence, the storage requirements for a 1-second simulation run are 16 times smaller per timestep when using an AR phase screen versus a translating phase screen. For a 4-second simulation, the analogous factor is 113.8$\times$ smaller. Memory requirements for the AR method are no different than those for phase screen extrusion methods: the screen size for each timestep is the determined by the desired aperture size and resolution. 

For computation comparisons, $N$log$_2(N)$ for the $N = 384^2$, $1536^2$ and $4096^2$ screen size cases is $2.5\times 10^6$, $5\times 10^7$ and $4\times 10^8$ respectively, which indicate the AR method is $20\times$ and $160\times$ more efficient for 1-second and 4-second simulations.

\subsection{Periodicity}
\label{ssec:periodicity}

Sample-based Fourier methods suffer from the problem of periodicity \cite{Mcglam76}, in that phase screens wrap around the domain. In the case of AR-generated phase screens, the addition of scaled noise at each timestep results in a substantially uncorrelated phase screen after a time interval dependent on $|\alpha|$. For example, after 500 timesteps, with $|\alpha|=0.99$, less than 1\% of the phase from the initial timestep is retained ($0.99^{500} = 0.0066$). The corresponding interval for $|\alpha|=0.999$ is 5000 timesteps. A phase screen for a given wind layer with velocity $\sim 8$ m/s would take approximately 1 second to traverse the Gemini aperture, hence using $|\alpha|=0.999$ for that layer would require a screen spanning a spatial extent $5\times$ larger than the aperture at an AO system rate of 1 KHz to minimize periodicity. However, this screen size would not change with wind direction or simulation length, unlike for the traditional frozen flow screen generation methods. It is only dependent on wind speed, AO system rate and the chosen value of $|\alpha|$ for a given layer. 

\section{Analysis and validation}
\label{sec:validation}  

The set of tests we performed to validate our AR atmospheres can be classified as {\bf a)} verification of the shape and distribution of spatial and temporal power spectra to show that the AR phase screen generation method conforms to theory and {\bf b)} comparisons in AO simulator performance between pure frozen flow atmospheres and AR atmospheres to quantify how changes in assumptions about the atmosphere model affect system performance metrics. In the case of GPI, a planet imager that seeks to occult a target star and enhance contrast ratio in the region immediately surrounding the star, one key metric is residual scattered light at low radii centered on the parent star, a region where there is a high likelihood of discovering planets. The LSST project has also observed material changes in PSF recovery from the addition of ``boiling'' via the AR method to their simulations which is causing them to revisit assumptions about their atmosphere model \cite{Schneider15}. Characterizing the changes in system response to injected stochastic content is a necessary step in the process of deciding whether replacing the existing model with another is worth the effort. 

In both tests, atmosphere datacubes (time series of 2-D phase screens) were sized slightly larger than the Gemini South primary pupil ($N_p=384$ with a pixel scale of 0.0224 meters pixel$^{-1}$) and with $|\alpha|=0.99$ or $|\alpha|=0.999$. For the power spectrum checks, we generated higher-resolution (in time) AR atmosphere datacubes with 8192 timesteps (representing a 5.46-second exposure at the GPI frame rate of 1.5 KHz). Comparative simulations were fed atmosphere realizations for a 1 second exposure when analyzing point spread function (PSF) differences between AR and frozen flow screens. Longer 22-second exposures at 1 KHz were used when comparing to GPI telemetry, which was collected in blocks with the same parameters. Three wind layers with differing velocities and directions derived from conditions observed at Cerro Pachon were tracked \cite{Tokovinin06}. 

\subsection{Spatial and temporal power spectral densities}
\label{ssec:psds}

A Blackman window was used to generate an unbiased, low-leakage periodogram with interval length set to 1024 samples for AR atmosphere phase screen time series of varying lengths in time and scaling parameter, $\alpha$. Spatial PSDs were calculated at each radial spatial wavenumber within the GPI aperture by exploiting Parseval's theorem and compared to theory \cite{Johan94}. Temporal PSDs were plotted over a frequency range [-512,512] Hz from the generated periodograms \cite{Poyneer07}.

   \begin{figure}[h!]
   \includegraphics[width=0.5\columnwidth]{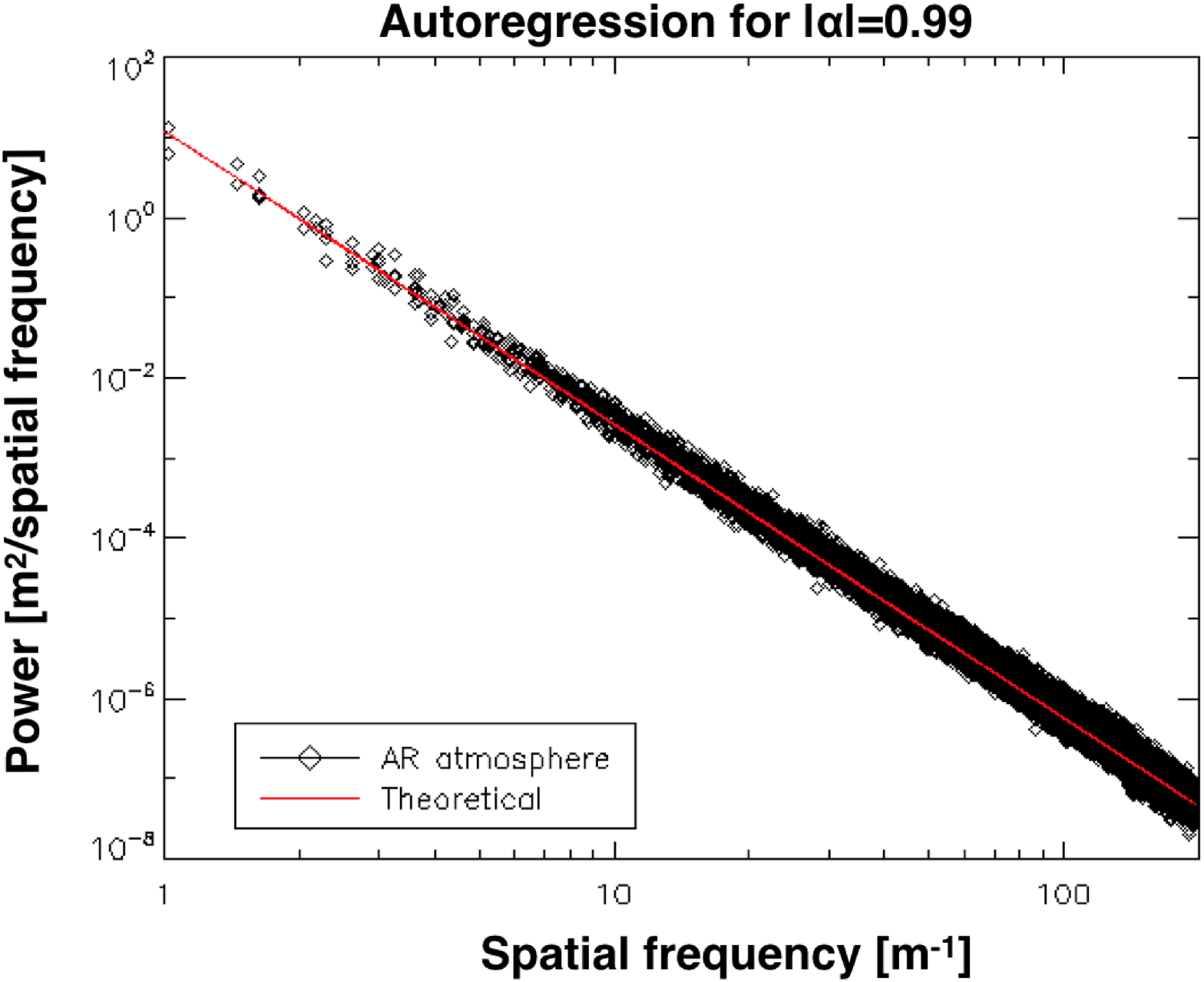} 
   \includegraphics[width=0.5\columnwidth]{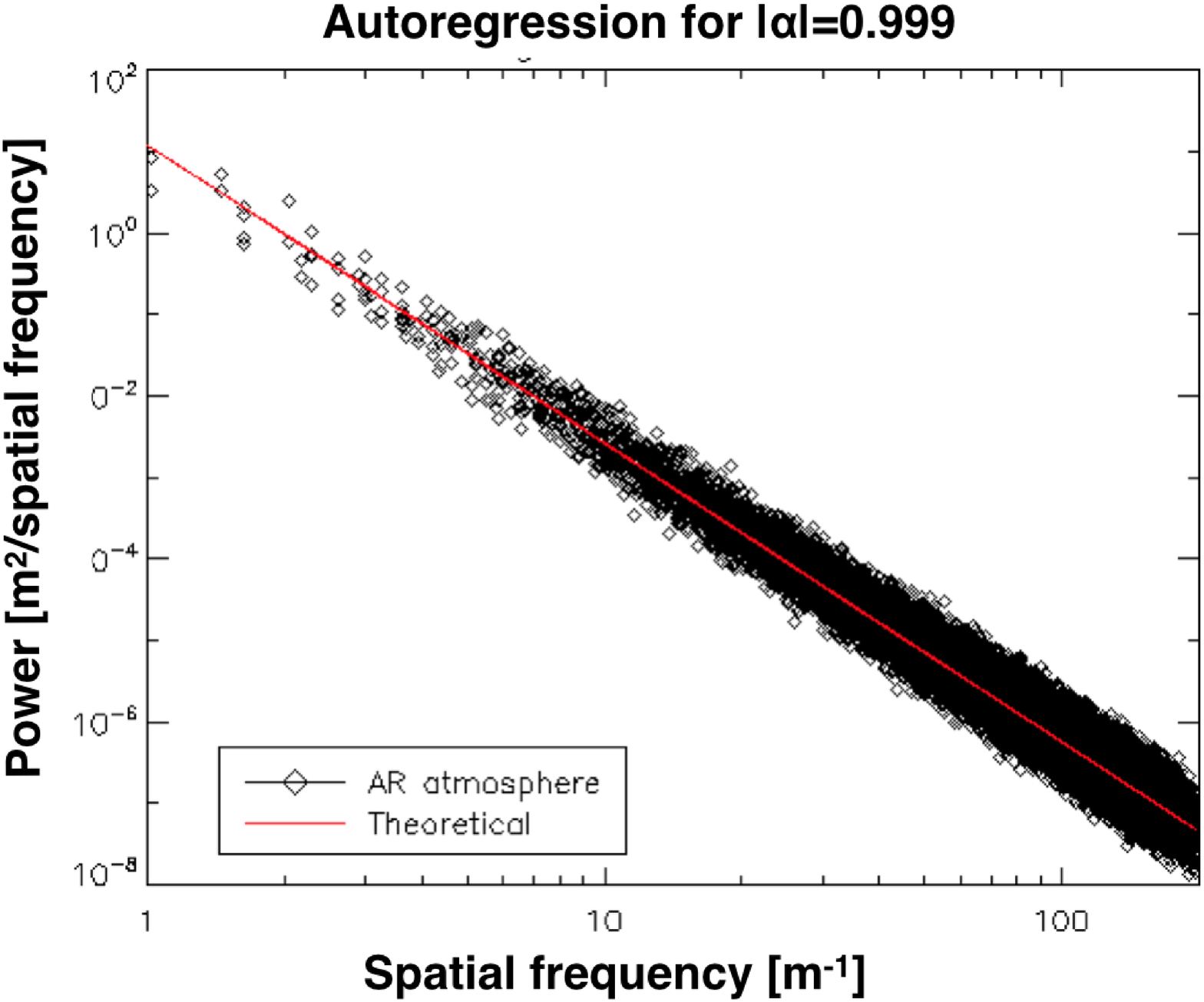}
   \caption
   { \label{fig:spat-psd} 
	Spatial PSDs for datacubes comprising time series of autoregressive phase screens with $|\alpha|=0.99$ (\textit{left}) and $|\alpha|=0.999$ (\textit{right}). Overplotted on both is the theoretical Kolmogorov power spectrum derived from \cite{Johan94}. Power spectral density of each datacube follows the expected  Kolmogorov slope ($\kappa^{-11/3}$). The $|\alpha|=0.99$ datacube exhibits lower variance about the theoretical slope because of the higher magnitude scaled noise injected into successive frames.}
   \end{figure} 



Spatial PSDs for $|\alpha|={0.99,0.999}$ are shown in Fig \ref{fig:spat-psd}. These atmosphere realizations used Kolmogorov statistics and demonstrate the expected $\kappa^{-11/3}$ slope for spatial frequency, $\kappa$ (the theoretical power spectrum is overplotted for comparison) \cite{Johan94}.  The PSD for $|\alpha|=0.999$ shows a greater variance around the theoretical slope because of the higher correlation between phase from one timestep to the next. The greater stochastic content in the $|\alpha|=0.99$ phase screens makes it hew closer to the power spectrum for infinite realizations of purely random screens. 

\begin{figure}[h!]
   \includegraphics[width=0.5\columnwidth]{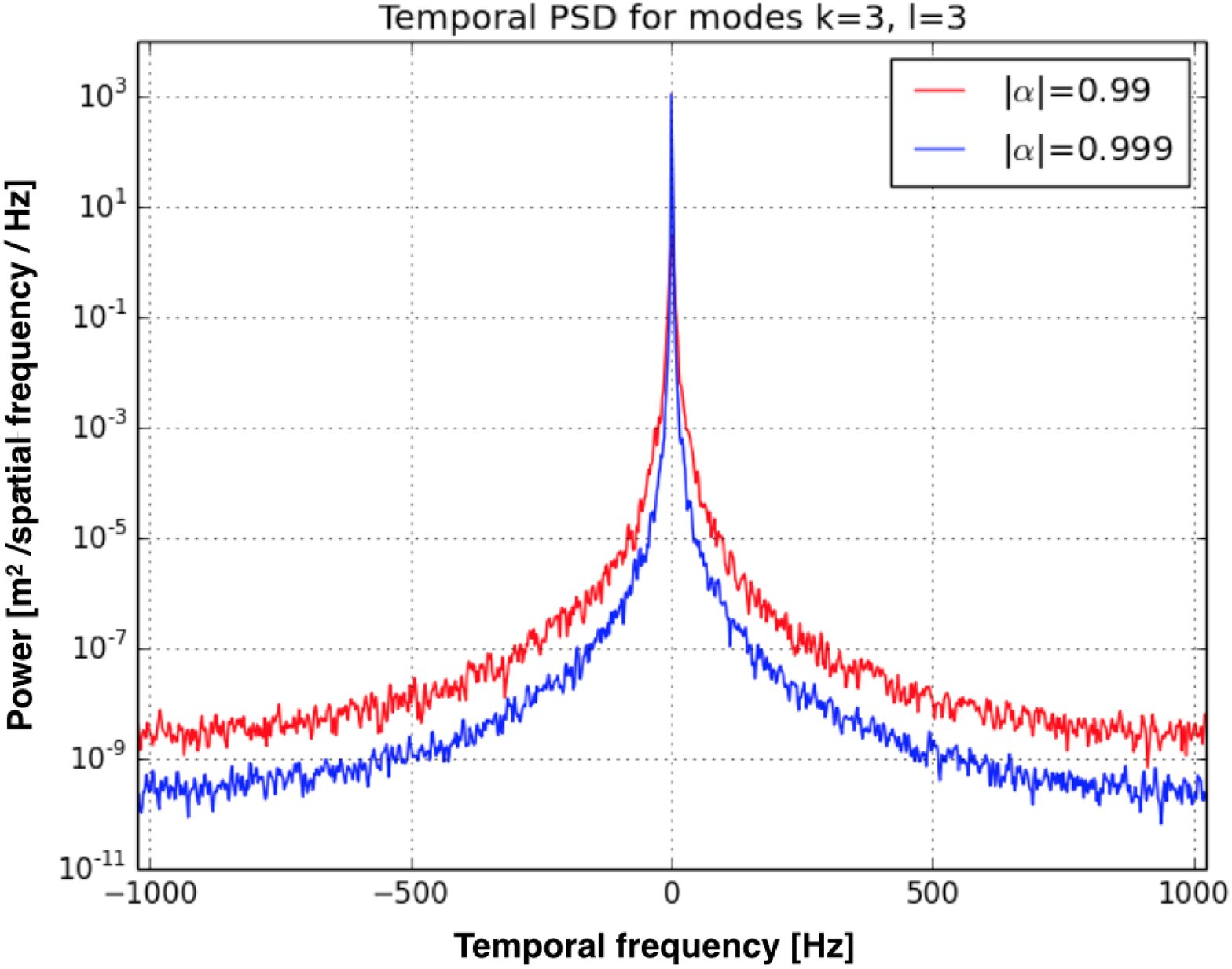}
   \includegraphics[width=0.5\columnwidth]{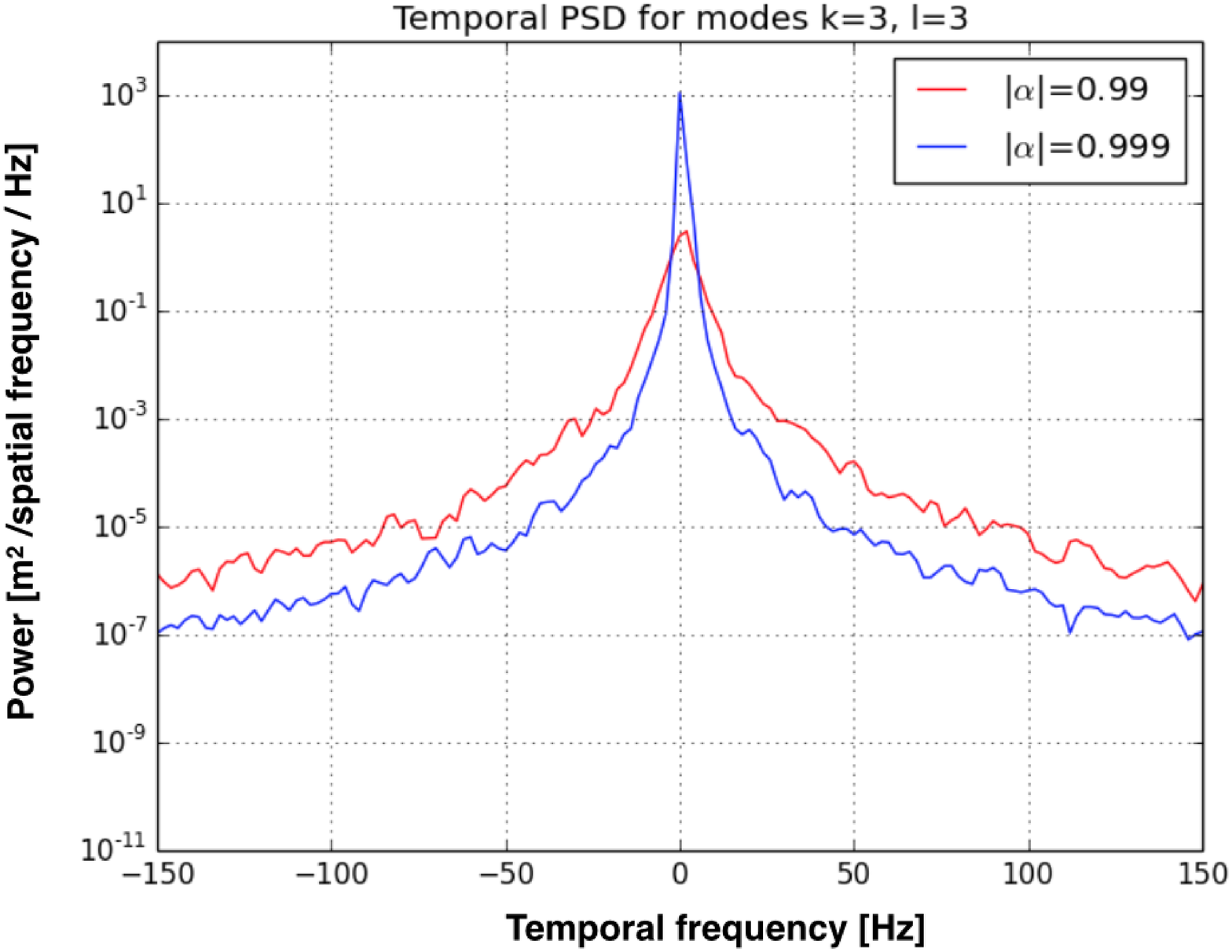}
   \caption
   { \label{fig:temp-psd} 
	Temporal power spectral density comparison for two sets of autoregressive phase screen time series with $|\alpha|=0.99$ and $|\alpha|=0.999$ plotted over the full temporal frequency range (\textit{left}) and zoomed in to lower frequencies (\textit{right}) to show the lower power peak for the $|\alpha|=0.99$ case. Integrated power is the same for both cases.}
   \end{figure}

Correctly scaling $\beta$ ensures overall power is conserved between datacubes with varying $\alpha$ -- Fig. \ref{fig:temp-psd} shows phase screen series with lower magnitudes of alpha have lower peaks and wider temporal power spectrum profiles, but the integral under the curve is constant as $|\alpha|$ varies. 

\subsection{Simulation tests}
\label{ssec:simtest}

AR and frozen flow atmosphere realizations were fed into the GPI AO simulator \cite{Poyneer06} and the spatial and temporal power content of the residual phase in each case was compared to test the sensitivity of modeled instrument behavior to atmosphere models with greater turbulence added. 

The GPI AO simulator is a detailed end-to-end system simulation with Fourier optics components. It uses a standard approach of translating large phase screens to generate a frozen flow atmosphere. The output of the AO system feeds an apodized-pupil Lyot coronagraph. The coronagraph suppresses diffraction and the spatial filter creates a dark hole where the residual scattered light is a function of both propagated wavefront sensor noise and residual atmosphere. Until the AR method was used for atmosphere generation, simulations were limited to 4-second exposure times, a constraint that has now been lifted. In addition, it is now possible to simulate aspects of GPI's calibration system, which compensates for slowly varying non-common path errors (like temperature, flexure and changing gravity vector) and operates on timescales of tens of seconds to minutes.


   \begin{figure}[h!]
   \includegraphics[width=1.0\columnwidth]{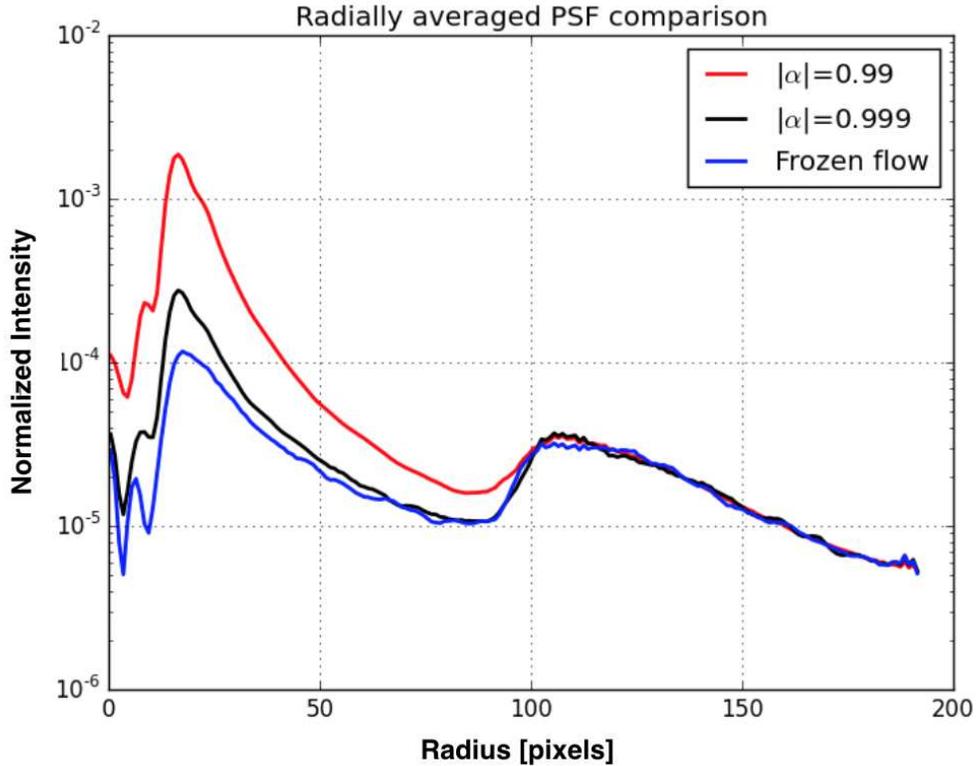}
   \caption
   { \label{fig:arVff} 
	Radially averaged profile of science PSF image produced by the GPI AO simulator running in closed-loop for three realizations of the atmosphere: pure frozen flow (blue), autoregressive with $|\alpha|=0.999$ (black) and $|\alpha|=0.99$ (red). Intensity on the y-axis is normalized such that the peak of the ideal, unblocked PSF is 1. Inside the innermost 10-pixel radius, PSF intensity is suppressed by coronagraphy. Outside the corrected area at a radius of $\sim100$ pixels, the power in all PSFs converges. At lower orders, the AR atmospheres have greater residual power: $> 2\times$ for $|\alpha|=0.999$ and $\sim 20\times$ for $|\alpha|=0.99$). The scattered light at a spatial location in the PSF corresponds directly to a specific Fourier mode in the spatial PSD.}
   \end{figure} 


We observed that the AR-generated atmospheres resulted in PSF images with more scattered light (higher intensity) at lower radii when compared to pure frozen flow ($2\times$ for $|\alpha|=0.999$ and $\sim 20\times$ for $|\alpha|=0.99$). Fig. \ref{fig:arVff} shows a PSF comparison of the default controller for each atmosphere type: pure frozen flow, AR with $|\alpha|=0.999$, and AR with $|\alpha|=0.99$. The plot is a radial average of the normalized intensity of the instantaneous science PSF recorded half way through a 1-second simulation run. Intensity is normalized such that the peak of the ideal, unblocked PSF is set equal to 1. Scattered light at a given location in the PSF plot corresponds to residual power for a specific Fourier mode \cite{Perrin03}. Hence, the excess power seen close to the center of the PSF (at low radii) translates into more power at low order Fourier modes, $[k,l]$. Below a radius of 10 pixels, power is suppressed by the apodized Lyot coronagraph. However, in the 10-100 pixel radius region (GPI's ``dark hole'' \cite{Poyneer14}), significantly different PSFs were recorded for small variations in atmospheric power distributions, which impacts the ability of the instrument to detect planets that lie close to the target star. To verify that this was a feature inherent in AR atmospheres that could be exploited when comparing to telemetry or a vestige of artefacts introduced by GPIAOS, we compared the temporal PSDs of the input atmosphere models.

\begin{figure*}
   \includegraphics[width=0.5\columnwidth]{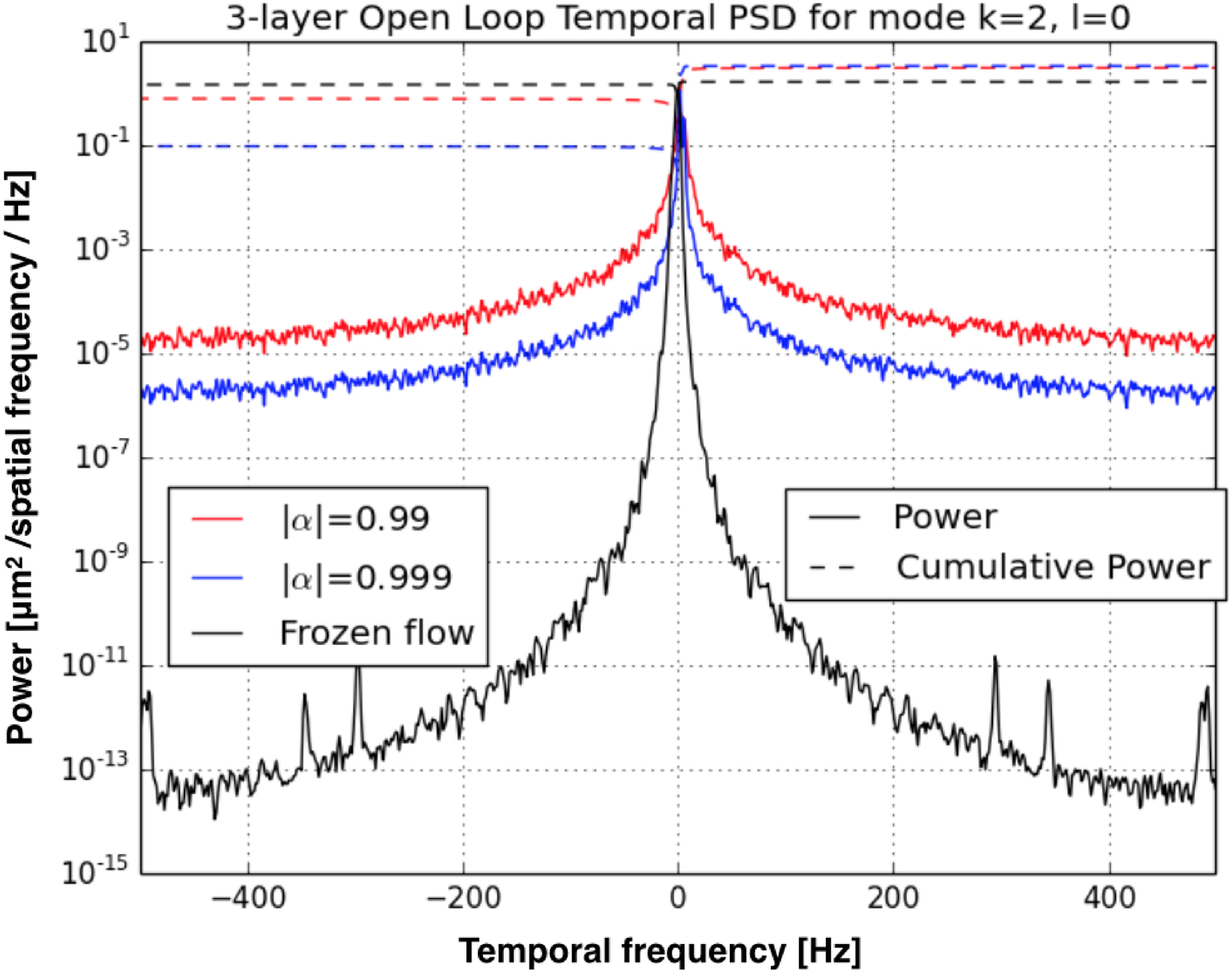}
   \includegraphics[width=0.5\columnwidth]{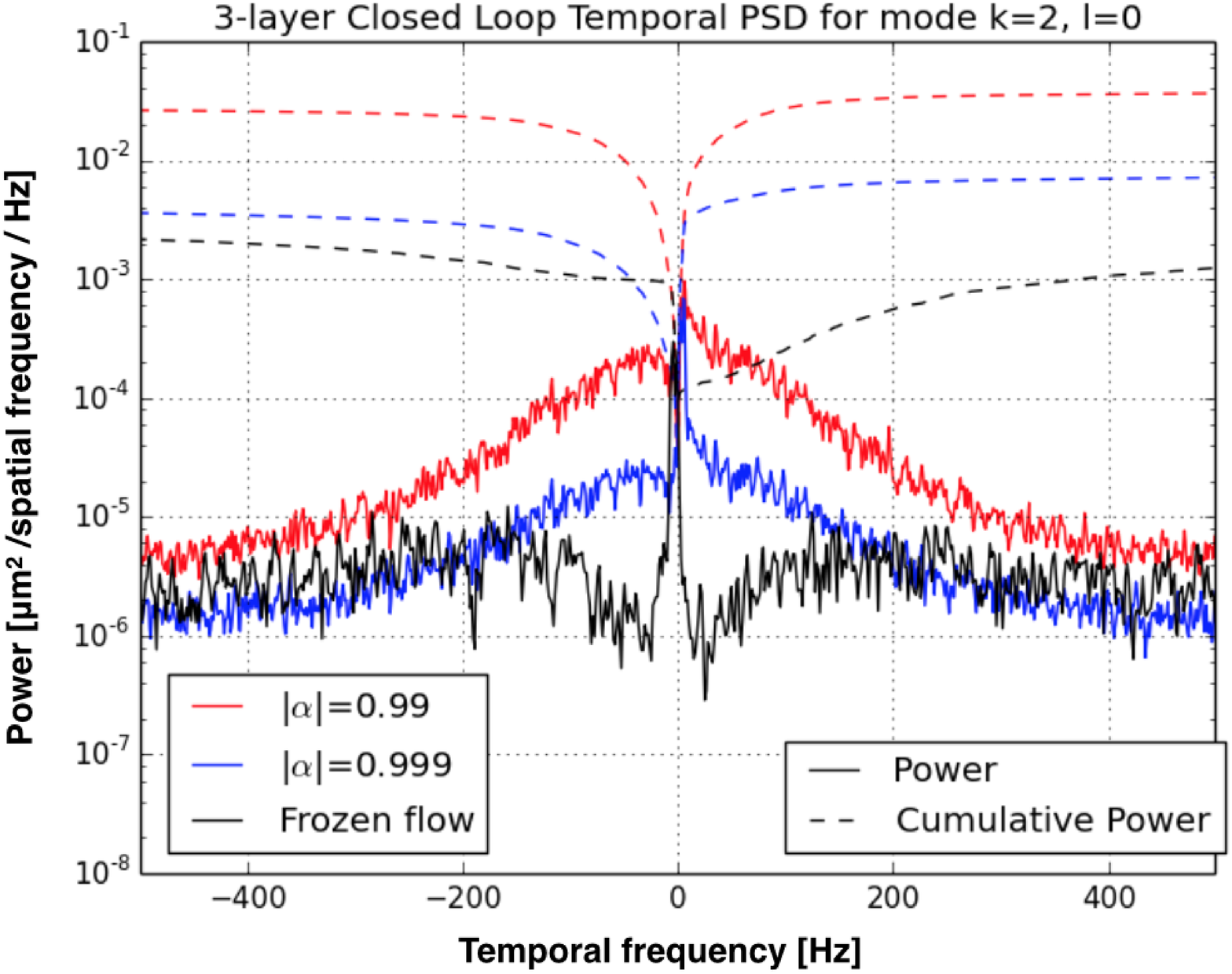}
   \caption
   { \label{fig:temppsd-olvcl} 
	Temporal power spectral comparison for open loop atmospheres (\textit{left}) versus closed loop residuals ({\textit{right}}) for AR atmospheres ($|\alpha|=0.99$ and $|\alpha|=0.999$, over a 22-second interval) and pure frozen flow (restricted to 4-second exposures by memory and computation limits).  The AR atmospheres have much broader peaks than the frozen flow atmosphere realization in lower order modes beyond 5 Hz in the left plot, hence rejection is worse and residuals exhibit more power in the right hand figure. This manifests itself as a brighter PSF core outside the central obscuration in a science image. Dashed lines represent cumulative power from $0$ Hz to $\pm 512$ Hz.}
   \end{figure*}

Analysis of the temporal power spectra of AR atmospheres versus pure frozen flow atmospheres reveals the source of the elevated low-order intensity in Fig. \ref{fig:arVff}. Fig. \ref{fig:temppsd-olvcl} plots the open- and closed-loop temporal PSDs, \textit{i.e.} the power content of the AR atmosphere datacubes (open-loop) and that of the residuals as seen by the system's wavefront sensor using the default controller (closed-loop). The two types of atmospheres have similar low-frequency power but the broad wings beyond 5 Hz in the open-loop AR PSD are not suppressed resulting in broad wings in the closed-loop PSD and, hence, more scattered light close to the center of the PSF in Fig. \ref{fig:arVff}. The residuals for an AR atmosphere with $|\alpha|=0.99$ in the right-hand plot in Fig. \ref{fig:temppsd-olvcl} show elevated power levels (the area under the curve is $\sim 20\times$ greater) over the frozen flow model atmosphere consistent with the differences seen in Fig. \ref{fig:arVff}.

The temporal structure of the atmosphere model evidently has a large impact on system performance. The magnitude of the impact can be estimated using the error transfer function (ETF) of a simulator because a change in input PSD changes the error output of a closed loop system \cite{Roddier}.  The ETF for GPIAOS  \cite{Poyneer06}, which applies corrections two frames after measurement, is given by:

\begin{equation}
ET\hspace{-0.5mm} F = \left| \frac{1}{1 + z^{-2} C(z)} \right|^2
\label{eqn:etf}
\end{equation}

\ni where:
\vspace{-2mm}
\begin{itemize} \itemsep1pt \parskip0pt \parsep0pt
\item $C(z) = g/(1 - c z^{-1})$, $0 < g < 1$ is the default modal gain and $c\lesssim 1$ is integrator leak gain.
\end{itemize}

The temporal PSD of the measurements of the AO system in closed loop is estimated by: 

\begin{equation}
P_{CL} = ET\hspace{-0.5mm} F(P_{\phi} + P_N) 
\label{eqn:psd_cl}
\end{equation}

\ni where:
\vspace{-2mm}
\begin{itemize} \itemsep1pt \parskip0pt \parsep0pt
\item $P_{\phi}$ is the temporal PSD of the phase (measured when the system applies no correction).
\item $P_N$ is the temporal PSD of the WFS noise (measured when the system is run closed-loop with no phase aberrations passed through it {\em i.e.} with no phase screen).
\end{itemize}

We can neglect the impact of the WFS response on the atmosphere as it is significant only at high temporal frequencies. Further details on this type of modeling are outlined in Ch. 6 of \cite{Roddier}.

More than an exercise in curiosity, in at least the GPI case, greater scattered light near the core of the PSF has the very real effect of decreasing contrast and reducing sensitivity to and detection of exoplanets. Hence, simulating the operation of the system with more realistic atmosphere models that have the same statistics as recorded telemetry will provide more credible limits to the system's capabilities. Using such models will allow tests of control schemes that improve contrast in conditions that more closely resemble the real-world observing environment than pure frozen flow atmosphere models. In the next section we describe our ongoing effort to exploit the flexibility and customizability of the AR method to build a better atmosphere model by extracting meaningful parameters like wind speed and direction, wind layers with the most power content, and statistics like $r_0$ and $L_0$. 

\section{Motivation}
\label{sec:motivation}

The magnitude of the effect of introduced ``boiling'' on PSF quality, plotted in Fig. \ref{fig:arVff}, illustrates the need to better understand atmospheric characteristics, find ways to measure them and correct for them. In the previous sections we have outlined an AR model with a number of degrees of freedom that account for and temporally evolve aspects of atmospheric turbulence such as statistics ($r_0$, $L_0$), ``boiling'', and wind layers with differing wind velocities and directions while still remaining computationally tractable and suitable for large-scale simulations. Tuning those degrees of freedom to improve upon prior models and better reflect reality by comparing to gathered telemetry is the next phase.

We are in the process of analyzing telemetry gathered by AO systems on GPI and the Shane telescope at Lick Observatory \cite{Gavel14}, by performing AR model fits to temporal PSDs and extracting parameters from spatial PSDs. This effort, in turn, feeds into implementing wind predictive AO \cite{Poyneer10} on a test bed \cite{Rudy15} and on-sky to eliminate temporal errors caused by wind-blown phase. The wind predictive method employs Kalman filters and Linear Quadratic Gaussian (LQG) control to notch out power in wind peaks found. The AR method will be used to generate synthetic atmospheres on which the Kalman filter generator will be trained. We wish to explore the sensitivity of Kalman filters to power in the random element of the atmosphere and crafting a realistic atmosphere model is the first step in the process. 

In simulations, for each realization of the atmosphere the GPI AO simulator was run with the default unoptimized controller, an optimized gain filter and a Kalman filter \cite{Poyneer07}. In all simulation comparisons, the Kalman filter had the greatest residual suppression (especially at higher orders), followed by the optimized filter. The Kalman filter achieved the least improvement for $|\alpha|=0.99$, which is expected behavior when a filter optimized for flow is applied to an atmosphere with higher power in boiling, but was still comparable to or better than an optimized filter. 

\begin{figure}
   \includegraphics[width=1.0\columnwidth]{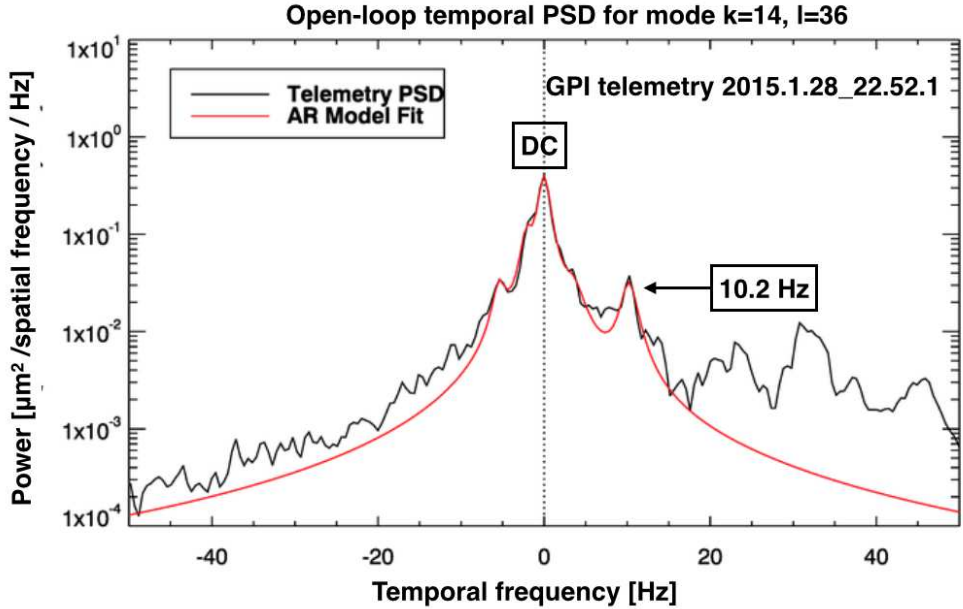}
   \caption{ 
   \label{fig:gpitelem} 
   Temporal PSD of Fourier mode $k=14, l=36$ for open-loop phase reconstructed from closed-loop 
   residuals as measured by the Gemini Planet Imager (solid black line) with the overlaid model
   fit (solid red line) for peaks at DC ($|\alpha|=0.995$) and 10.2 Hz ($|\alpha|=0.993$).}
   \end{figure}

One example of peak fitter and layer finder output is shown in Fig. \ref{fig:gpitelem}. Closed loop residual phase as measured by GPI's wavefront sensor over 22.1 seconds with the system pointed at a bright star and an AO system rate of 1 kHz was stored at the date and time indicated on the plot. Using the system's error and noise transfer functions, pseudo-open loop phase is estimated, which is then converted to Fourier modes and temporal PSDs calculated. A PSD peak fitter and layer finder algorithm \cite{Poyneer09} identifies and fits wind peaks and then finds wind layers. The width of each peak sets the magnitude of $\alpha$ for that peak; wider peaks have lower values of $|\alpha|$, as seen in Fig. \ref{fig:temp-psd}. The temporal frequency of the peak is used to set the phase of $\alpha$ per equation \ref{eqn:alphaphase}. Fig. \ref{fig:gpitelem} plots the temporal PSD of a Fourier mode ($k=14, l=36$) displaying AR model fits to power peaks at 0 Hz (DC, $|\alpha|=0.995$) and 10.2 Hz ($|\alpha|=0.993$). Values of $|\alpha|$ for the other peaks fit range from 0.991 to 0.996, a characteristic shared by other telemetry sets analyzed which motivates our restriction of $|\alpha|$ values analyzed in AR modeling to the interval $[0.990, 0.999]$. Two wind layers found have wind velocities of 1.2 m/s and 4.9 m/s (with no altitude information).

Only wind peaks which contain significant power (defined to be greater than 20\% of the peak with the highest power content, the D.C. peak in this case) are identified and used for wind prediction, {\em i.e.} ones that are worthwhile for the system to correct. Hence, peaks at -2, -5 and beyond +20 Hz are ignored. 

In contrast to the method outlined in \cite{Glind93}, we have observed no frequency dependence of the decorrelation parameter $\alpha$. Our fits to telemetry indicate that from one time step to the next its magnitude is confined to the interval $[0.990, 0.999]$ which is significantly different from the implied range of $[1/\sqrt{3},1/\sqrt{2}]$ in the prior work. 

The process of analyzing telemetry is work in progress, but there is already evidence to suggest that the AR model is a better representation of the power distribution of the atmosphere (between frozen flow and ``boiling'' components) than pure Kolmogorov frozen flow. Long simulations also need information about the temporal evolution of atmospheric characteristics like power in wind layers, $r_0$ and $L_0$ (both of which are extracted from the spatial PSD of atmospheric telemetry). Telemetry collection through individual nights in varied conditions from different sites and its analysis will better inform how to set and evolve atmosphere model parameters in long exposure simulations. 

\section{Conclusions and future work}
We have implemented, validated, tested and analyzed an autoregressive method for the generation of phase screens with turbulence. Where comparisons can be made, the AR atmosphere technique is 10-160 times more efficient in its use of computational resources and memory than sample-based phase translation methods. By raising the limit on simulation length, it significantly extends the capabilities of existing simulators and allows the testing of longer time-scale corrections. The method's flexibility allows for the easy incorporation of any desired atmospheric turbulence model and and variation of parameters (like coherence length and outer scale). Finally, its simplicity when compared to other worthy attempts at optimization makes it ready for immediate use and adaptation in simulation (source code in the form of IDL and python routines is available on \verb+github+). 

Spatial power spectra are as expected (and can be tweaked based on the turbulence model used -- we chose Kolmogorov for this exercise). Temporal power spectra show AR atmospheres have very different power distributions compared to frozen flow realizations which results in significant differences in PSF profiles in simulations. Work is ongoing to relate AR atmospheres to real-world telemetry gathered by multiple AO systems so that the statistics of generated atmospheres (\textit{e.g.} power content, turbulence strength and evolution of parameters like $r_0$) are as close as possible as those extracted from observations. Tailored atmosphere models that resemble real conditions will then be used to train Kalman filter generators for optimizing Fourier Wind Identification schemes before they are tested on-sky.

\section{Acknowledgment}
The work of SS and ARR is funded by the UC Lab Fees Research Program grant 12-LF-236852.

This work is performed under the auspices of the U.S. Department of Energy 
by Lawrence Livermore National Laboratory under Contract DE-AC52-07NA27344.
The document number is LLNL-JRNL-667773.

SS wishes to thank Erik Johansson at the NSO for his insight during highly productive 
discussions about the AR method and Jim Chiang at SLAC for the considerable assistance 
he provided in creating the python version of the AR screen generation code.

We thank the anonymous referees for their constructive and insightful criticism and 
commentary.

The Gemini Observatory is operated by the 
Association of Universities for Research in Astronomy, Inc., under a cooperative agreement 
with the NSF on behalf of the Gemini partnership: the National Science Foundation 
(United States), the National Research Council (Canada), CONICYT (Chile), the Australian 
Research Council (Australia), Minist\'{e}rio da Ci\^{e}ncia, Tecnologia e Inova\c{c}\~{a}o 
(Brazil) and Ministerio de Ciencia, Tecnolog\'{i}a y Innovaci\'{o}n Productiva (Argentina)


\begin{thebibliography}{99}


\bibitem{Rudy14}
Rudy, A.~R., Srinath, S., Poyneer, L.~A., Ammons, S.~M., Gavel, D.~T., Kupke,
  R., Dillon, D., and Rockosi, C., ``Progress towards wind predictive control
  on ShaneAO: Test bench results,'' \pspie {\bf 9148}, 91481Z (2014).

\bibitem{Poyneer06}
Poyneer, L.~A. and Macintosh, B.~A., ``Optimal Fourier control performance
  and speckle behavior in high-contrast imaging with adaptive optics,'' 
  \opex {\bf 14}, 7499--7514 (2006).

\bibitem{Connolly10}
Connolly, A.~J., Peterson, J., Jernigan, J.~G., Abel, R., J.~Bankert, C.~C.,
  Claver, C.~F., Gibson, R., Gilmore, D.~K., Grace, E., Jones, R.~L., Ivezic,
  Z., Jee, J., Juric, M., Kahn, S.~M., Krabbendam, V.~L., Krughoff, S., Lorenz,
  S., Pizagno, J., Rasmussen, A., Todd, N., Tyson, J.~A., and Young, M.,
  ``Simulating the LSST system,''  \pspie {\bf 7738}, 77381O (2010).

\bibitem{Wang12}
Wang, L. and Ellerbroek, B., ``Computer simulations and real-time control of
  ELT AO systems using graphical processing units,'' \pspie {\bf 8447}, 
  844723 (2012).

\bibitem{Jee11}
Jee, J.~M. and Tyson, J.~A., ``Towards precision LSST weak-lensing measurement- I: impacts of
atmospheric turbulence and optical aberration'', Pub. Astr. Soc. Pacific {\bf 123}, 596--614 (2011)

\bibitem{Mcglam76}
McGlamery, B., ``Computer simulation studies of compensation of turbulence
degraded images,''  \pspie {\bf 74}, 225--233 (1976).

\bibitem{Peterson15}
Peterson, J.~R., Jernigan, J.~G., Kahn, S.~M., Rasmussen, A.~P., Peng, E., Ahmad, Z., Bankert, J., Chang,  
 C., Claver, C., Gilmore, D.~K., Grace, E., Hannel, M., Hodge, M., Lorenz, S., Lupu, A., Meert, A.,  
 Nagarajan, S.,  Todd, N., Winans, A. and Young, M.,
 ``Simulation of astronomical images from optical survey telescopes using a comprehensive photon 
   Monte Carlo approach,'' \apj Suppl. {\bf 218}, 14 (2015)

\bibitem{Assemat06}
Ass{\'{e}}mat, F., Wilson, R.~W., and Gendron, E., ``Method for simulating
  infinitely long and non stationary phase screens with optimized memory
  storage,'' \opex {\bf 14}, 988--999 (2006).

\bibitem{Fried08}
Fried, D.~L. and Clark, T., ``Extruding Kolmogorov-type phase screen ribbons'', \josaa {\bf 25},
463--468 (2008)

\bibitem{Beghi13}
Beghi, A., Cenedese, A., and Masiero, A., ``Multiscale phase screen synthesis based on local principal component analysis'', \ao {\bf 52}, 7987--8000 (2013)

\bibitem{Vorontsov08}
Vorontsov, A.~M., Paramonov, P.~.V., and Valley, M., ``Generation of infinitely-long phase screens for
modeling optical wave propagation through turbulence,'' \pspie {\bf 5891},  589108 (2005).

\bibitem{Johan94}
Johansson, E.~M. and Gavel, D.~T., ``Simulation of stellar speckle imaging,''
 \pspie {\bf 2200}, 372--383 (1994).

\bibitem{Harding03}
Harding, C.~M., Johnston, R.~A., and Lane, R.~G., ``Fast simulation of a Kolmogorov phase screen,''
\ao {\bf 38}, 2161--2170 (2003).

\bibitem{Glind93} 
Glindemann, A., Lane, R.~G., and Dainty, J.~C., ``Simulation of time-evolving speckle patterns using
Kolmogorov statistics,'' \jmo {\bf 40(12)}, 2381--2388 (1993).

\bibitem{Welsh97}
Welsh, B.~M., ``A Fourier series based atmospheric phase screen generator for simulating anisoplanatic geometries and temporal evolution,'' \pspie {\bf 3125},  327--338 (1997).

\bibitem{Lane92}
Lane, R.~G., Glindemann, A., and Dainty, J.~C., ``Simulation of a Kolmogorov
phase screen,'' Waves in Random Media {\bf 2}, 209--224 (1992).

\bibitem{Sedmak04}
Sedmak, G., ``Implementation of fast-Fourier-transform-based simulations of extra-large atmospheric
phase and scintillation screens,'' \ao {\bf 43}, 4527--4538 (2004)

\bibitem{Gavel14}
Gavel, D., Kupke, R., Dillon, D., Norton, A., Ratliff, C., Cabak, J., Phillips, A., Rockosi, C., Mcgurk, R., Srinath, S., Peck, M., Lanclos, K., Gates, J., Saylor, M., Ward, J., and Pfister, T. ``ShaneAO: wide science spectrum adaptive optics system for the Lick Observatory,'' \pspie {\bf 9148}, 914805 (2014) 

\bibitem{Percival93}
Percival, D.~B., and Walden, A.~T., {\em Spectral Analysis for Physical Applications} (Cambridge University, reprint 1998)

\bibitem{Poyneer07}
Poyneer, L.~A., Macintosh, B.~A., and V\'eran, J.-P., ``Fourier transform
  wavefront control with adaptive prediction of the atmosphere,'' \josaa 
  {\bf 24},  2645--2660 (2007).

\bibitem{Fried65}
Fried, D.~L., ``Statistics of a geometric representation of wavefront distortion,'' \josaa 
{\bf 55}, 1427-–1435 (1965)

\bibitem{Frigo05}
Frigo, M., and Johnson, S.~G., ``The design and implementation of FFTW3,'' 
  Proc. IEEE {\bf 93}, 216-–231 (2005).

\bibitem{Richards07}
Richards, M.~A., ``The discrete-time Fourier transform and discrete Fourier transform of windowed
stationary white noise,'' \texttt{http://users.ece.gatech.edu/mrichard/DFT\%20of\%20Noise.pdf}

\bibitem{Tokovinin06}
Tokovinin, T. and Travouillon, T., ``Model of optical turbulence profile at Cerro Pachon'',
MNRAS {\bf 365}, 1235-1242 (2006)

\bibitem{Schneider15}
Schneider, M~D., Lawrence Livermore National Lab, Livermore, CA, USA (personal communication, 2015)

\bibitem{Perrin03}
Perrin, M.~D., Sivaramakrishnan, A., Makidon, R.~B., Oppenheimer, B.~R., and Graham, J., ``The structure
of high Strehl ratio point-spread functions,'' \apj {\bf 596}, 702--712 (2003). 

\bibitem{Poyneer14}
Poyneer, L.~A., De~Rosa, R.~J., Macintosh, B., Palmer, D.~W.,
   Perrin, M.~D., Sadakuni, N., Savransky, D., Bauman, B., 
  Cardwell, A., Chilcote, J.~K., Dillon, D., Gavel, D., 
  Goodsell, S.~J., Hartung, M., Hibon, P., Rantakyrö, F.~T., 
  Thomas, S., and Veran, J-P.,
 ``On-sky performance during verification and commissioning of the
  Gemini Planet Imager's adaptive optics system,'' \pspie {\bf 9148},  
  91480K (2014).

\bibitem{Roddier}
Roddier, F., {\em Adaptive Optics in Astronomy} (Cambridge University, 1999)

\bibitem{Poyneer10}
Poyneer, L.~A. and Veran, J-P., ``Kalman filtering to suppress spurious signals in adaptive optics control,'' \josaa {\bf 27}, A223--A224 (2010).

\bibitem{Rudy15}
Rudy, A.~R., Poyneer, L.~A., Srinath S., Ammons, S.~M. and Gavel D., ``A laboratory demonstration of an LQG technique for correcting frozen flow turbulence in adaptive optics systems'', in prep

\bibitem{Poyneer09}
Poyneer, L.~A., {van Dam}, M.~A. and Veran, J-P. "Experimental
  verification of the frozen flow atmospheric turbulence assumption with use of
  astronomical adaptive optics telemetry," \josaa {\bf 26}, 833--846 (2009)
  
\end{thebibliography}
\end{document}